# Divergent Perspectives on Expert Disagreement: Preliminary Evidence from Climate Science, Climate Policy, Astrophysics, and Public Opinion[*]

James R. Beebe (University at Buffalo), Maria Baghramian (University College Dublin), Luke Drury (Dublin Institute for Advanced Studies), Finnur Dellsén (Inland Norway University of Applied Sciences)

We report the results of an exploratory study that examines the reactions of climate scientists, climate policy experts, astrophysicists, and non-experts ($N = 3,367$) to instances of disagreement within climate science and astrophysics. The study explores respondents' judgments about the factors that contribute to the creation and persistence of those disagreements and how one should respond to disagreements among experts. We found that, as compared to educated non-experts, climate experts believe (i) that there is less disagreement within climate science about climate change, (ii) that more of the disagreement that does exist concerns public policy questions rather than the science itself, (iii) that methodological factors play less of a role in generating existing disagreement among experts about climate science, (iv) that fewer personal and institutional biases influence the nature and direction of climate science research, (v) that there is more agreement among scientists about which methods or theoretical perspectives should be used to examine and explain the relevant phenomena, (vi) that disagreements about climate change should not lead people to conclude that the scientific methods being employed today are unreliable or incapable of revealing the truth, and (vii) that climate science is more settled than ideological pundits would have us believe and settled enough to base public policy on it. In addition, we observed that the uniquely American political context predicted participants' judgments about many of these factors. We also found that, commensurate with the greater inherent uncertainty and data lacunae in their field, astrophysicists working on cosmic rays were generally more willing to acknowledge expert disagreement, more open to the idea that a set of data can have multiple valid interpretations, and generally less quick to dismiss someone articulating a non-standard view as non-expert, than climate scientists.

[*] This article represents one of the central research outputs of the When Experts Disagree Project, which was funded by an Irish Research Council New Horizons Strand Two Interdisciplinary Award (Nov., 2015 – Nov., 2017).

## 1. Introduction

In today's complex societies, many policy decisions depend crucially upon expert advice and opinion. But experts can and do disagree, sometimes vehemently, and not all their disagreements are easily adjudicated. An immediate question facing policy makers, and in particular those involved in decisions concerning some of the greatest challenges facing humanity, such as environmental policy, is how to react to persistent disagreement among experts. A further issue concerns how to respond to the sorry corollary of scientific disagreements, viz., the frequent misrepresentation and misunderstanding of them in the media and civic society.

The current paper is the output of an interdisciplinary investigation, involving scientists and philosophers, of the ill understood, but socially and politically significant phenomenon of expert peer disagreement. The ultimate goal of the project is to gain a better understanding of the role and consequences of disagreement among scientific experts and its impact on policy decisions by governmental agencies and the formation of public opinion. We examined cases of expert scientific disagreement in a field that is relatively free of economic and political pressures (cosmic ray physics) and contrasted it with expert disagreements in a field where significant economic and political interests are at stake (climate science).[1]

Existing studies of the opinions of climate scientists (Oreskes 2004, Bray & von Storch 2008, Anderegg et al. 2010, Cook et al. 2013, Stenhouse et al. 2014, Cook et al. 2016) and the general public (Nisbet & Myers 2007, Leiserowitz et al. 2010, Leiserowitz et al. 2012, Pew 2012, Leiserowitz et al. 2015, Pew 2016, Gallup 2017) have shown significant disparities in their assessments of the extent of disagreement or consensus within climate science. These studies have not, however, examined the views of climate experts and non-experts in regard to the various personal, methodological, or institutional factors that these individuals think generate expert disagreement or how they think such disagreements should be approached. These studies also do not compare disagreements about climate science to disagreements in other scientific domains. Thus, the central questions we investigated included the following:

> (D1) In the politically charged discipline of climate science, are there significant differences between the opinions of climate scientists, climate policy experts, and non-experts regarding the extent and causes of disagreement within that discipline?
>
> (D2) Are there significant differences between the opinions of climate scientists, climate policy experts, and non-experts about the contributions made by what we call *"epistemic"* factors—i.e., issues that concern the quality of the data, reliability of available research methods, and familiar forms of scientific uncertainty—and *"nonepistemic"* factors—those that involve bias or ideological distortions of the ideal scientific process by outside influences in bringing about persistent disagreement?
>
> (D3) What do climate scientists, climate policy experts, and astrophysicists take to be the most appropriate reaction to scientific disagreements? In particular, (a) should persistent disagreement regarding a given theory decrease our confidence in that theory? And (b) does persistent

>disagreement lead lay people to have doubts about the possibility of objectivity in the domain?

---

[1] Cf. the online supplementary materials document that accompanies this article for a summary of the key disagreements among experts in cosmic ray physics.

(D4) What effects does scientific disagreement have on the field in question? In particular, are disagreements perceived as good for the health of the discipline itself and what effects does publicizing disagreement have on public trust in science?

We examined these issues using two questionnaires. We observed significant between-group differences in perceptions of the extent and causes of disagreement in climate science.

Our research questions and interpretations are informed by recent discussions within philosophy regarding disagreement between equally competent and otherwise equally well-informed agents—so-called 'peer disagreements' (Christensen 2007, 2009; Feldman 2009; Goldman 2009; Kelly 2005, 2010; Matheson 2014). A central issue at stake in cases of peer disagreement is the normative question of how someone should respond when she realizes that she is in such a situation, i.e., when she discovers that a peer disagrees with her.[2] Of the groups we studied, all except the astrophysicists were disinclined to move closer to the opinion of their peers in situations of this sort. We found that climate scientists were significantly more steadfast than their colleagues in astrophysics—at least with regard to disagreements about climate science. Relatedly, we found that climate scientists, compared with astrophysicists working on cosmic rays, are more inclined to doubt that experts who disagree with the dominant theoretical perspectives are as well informed as they are, which suggests that climate scientists are less trusting of apparent experts than their counterparts in astrophysics. This should perhaps not be surprising given the prevalence of ideological pundits masquerading as climate science experts in the media and elsewhere. The inherent uncertainty and data lacunae in the field of cosmic rays may also explain the willingness of astrophysicists working on cosmic rays to show greater tolerance towards non-standard views.

---

[2] Cf. the supplementary materials document for additional details about the philosophical debate.

**2. Research Materials**

In order to investigate the research questions described above, the When Experts Disagree project hosted workshops with astrophysicists, climate scientists, climate policy experts, and philosophers working on the epistemology of disagreement. At these workshops, we discussed our central research questions with each group and solicited their input on how we might investigate them in questionnaires we subsequently designed. On the basis of these discussions, we constructed two questionnaires, the first of which (Questionnaire 1) focused on descriptive questions about factors that lead to persistent expert disagreement. The second (Questionnaire 2) focused on normative questions such as the reasonability of expert disagreement and whether encounters with peers with whom one disagrees should make one less confident in one's opinions. Questionnaire 2 also investigated the implications of persistent disagreement for public policy decisions, public trust in science, and assessments of the reliability of the scientific field in question.

Our goal was to construct questionnaires for each participant group that were as similar as possible, so that comparisons across groups would be possible. However, some differences in wording were necessary. These differences are noted with underlining, parentheses, and footnotes below. Climate scientists, climate policy experts, and the comparison group we call "educated non-experts", consisting of university students and alumni, were all asked about disagreement among climate science experts. Our questions probed participants' judgments about climate change and climate science as a whole, as opposed to particular areas or issues within climate science.[3] Astrophysicists who work on cosmic ray physics were asked only about disagreements within cosmic ray physics. The items featured in Questionnaire 1 appear in Table 1.

> **We are interested in how <u>climate scientists</u> (astrophysicists) view disagreements about climate change (cosmic rays).[4]**
> 1.1 How much disagreement about <u>climate change</u> (cosmic ray physics) is there among the experts in your field?[5]
>
> **When considering the scientific methods and practices of your field, how important do you think the following factors are in the creation and persistence of disagreement about <u>climate change</u> (cosmic ray physics)?**
> 1.2 Those involved in the disagreement are not equally well informed.
> 1.3 Those involved in the disagreement begin from different starting points, prior assumptions, or theoretically motivated expectations.
> 1.4 Those involved in the disagreement highlight or focus on different kinds of data as evidence.
> 1.5 Those involved in the disagreement have differing views of the appropriate scientific method.
> 1.6 The issues about which experts in my field disagree are very complex.
> 1.7 The data about which experts in my field disagree involve a great deal of uncertainty.
> 1.8 It is difficult to obtain enough of the right kind of data needed to resolve the disagreements that arise.

---

[3] Some climate experts indicated that they would have liked to have commented on different aspects of the climate debate in different ways. However, because many members of the general public are unfamiliar with different aspects of the science of climate change and because we wanted to examine between-group differences on the same questions, we pitched our questions in this study at a general level. In a subsequent study, the results of which we do not report here, we asked different participant groups about particular issues within the overall climate debate.

[4] For climate policy experts, we added 'and climate policy experts' to 'climate scientists.' The heading for educated non-experts read "We are interested in how members of the general public view disagreements about climate change among climate science experts."

[5] Throughout both questionnaires, the phrases 'your field' and 'my field' were changed to 'climate science' for educated non-experts.

> **When considering the practices of other experts in your field, how important do you think the following factors are in the creation and persistence of disagreement about <u>climate change</u> (cosmic ray physics)?**
> 1.9 Those involved in the disagreement are motivated by political ideology to defend particular theories.
> 1.10 Those involved in the disagreement are motivated by financial incentives to defend particular theories.
> 1.11 Those involved in the disagreement defend certain theories because those theories represent their life's work and they cannot bear to give them up.
> 1.12 Those involved in the disagreement are trying to garner attention or make a name for themselves.
> 1.13 Those involved in the disagreement are simply being stubborn, closed-minded, or unreasonable.

*Table 1*. Elements of Questionnaire 1.

The answer choices for Question 1.1 were 'None,' 'Very little,' 'Some,' and 'A great deal.' For purposes of analysis, these were scored as 0, 1, 2, and 3. The answer choices for Questions 1.2 through 1.13 were 'Not important,' 'Slightly important,' 'Moderately important,' 'Important,' and 'Very important' (scored as 0, 1, 2, 3, and 4). All participants were asked demographic questions about their age, sex, education, and ethnicity or nationality. Experts were asked to describe the nature of their training and expertise.

In light of existing public opinion research (Nisbet & Myers 2007, Leiserowitz et al. 2010, Leiserowitz et al. 2012, Pew 2012, Leiserowitz et al. 2015, Pew 2016, Gallup 2017), we hypothesized that experts and lay people would make divergent judgments about how much disagreement there was within climate science. Given the political climate in the United States, we also expected that non-expert respondents in the U.S. would think there was more disagreement within climate science than their peers in other countries. Since scientific disagreements are rarely publicized except when connected to ideological, political, or religious disputes, we hypothesized that the non-experts would think that debates about climate science were driven more by ideological, social and personal factors than scientists in that field do.

The items featured on Questionnaire 2 appear in Table 2.

> **We are interested in how <u>climate scientists</u> (astrophysicists) view disagreements about <u>climate change</u> (cosmic rays).**
> 2.1 How much disagreement about <u>climate change</u> (cosmic ray physics) is there among the experts in your field?
>
> **Please indicate the extent to which you agree or disagree with the following claims:**
> 2.2 Two experts who are equally well-informed about the science of <u>climate change</u> (cosmic ray physics) might look at the same data but reasonably arrive at different conclusions.
> 2.3 Persistent disagreement about <u>climate change</u> (cosmic ray physics) may indicate that the tools or methods scientists use to study this phenomenon are not sufficiently reliable.
> 2.4 Persistent disagreement about <u>climate change</u> (cosmic ray physics) may mean that there is no correct theory in this domain.
> 2.5 Persistent disagreement about <u>climate change</u> (cosmic ray physics) may mean that there is more than one correct theory in this domain.
> 2.6 When I find that other experts who are as well-informed as I am hold opinions about <u>climate change</u> (cosmic ray physics) that are significantly different from my own, this sometimes leads me to become less confident in my own opinions.[6]
> 2.7 When I encounter an expert who disagrees with the dominant paradigms or theoretical perspectives in my field, this usually causes me to wonder whether they are really as well informed as other experts in the field.
> 2.8 The peer review process in my field is biased against publishing controversial hypotheses about <u>climate change</u> (cosmic ray physics).[7]
> 2.9 Minority or dissenting perspectives on <u>climate change</u> (cosmic ray physics) are often inappropriately silenced or suppressed within my field.[8]

---

[6] For educated non-experts, 'other experts' was changed to 'other people.'
[7] Educated non-experts were given "The peer review process in climate science is probably biased against publishing controversial hypotheses about climate change."
[8] Educated non-experts were given "Minority or dissenting perspectives on climate change are probably often inappropriately silenced or suppressed within climate science."

> 2.10 Disagreements about issues such as climate change (cosmic ray physics) can be good for the health of my field.
> 2.11 When experts in my field disagree about climate change, it is most often about which public policy recommendations should be made in light of the science rather than about the science itself.
> 2.12 Publicizing the extent of disagreement among experts in my field reduces public trust in science.
> 2.13 Scientific experts should present their science to the general public without making any policy recommendations about what society should do in light of the science.
> 2.14 The science of climate change is settled enough to base public policy on it.
> 2.15 How much trust do you have in climate science experts?
> 2.16 How much trust do you have in scientists working in other areas of science, e.g. astrophysics?

*Table 2*. Elements of Questionnaire 2.

Question 2.1 was identical to Question 1.1, so that we could compare participants' answers to other questions with their answers to this particular question on both questionnaires. The response choices for Questions 2.2 through 2.14 were 'Completely disagree,' 'Mostly disagree,' 'Slightly disagree,' 'Neither agree nor disagree,' 'Slightly agree,' 'Mostly agree,' and 'Completely agree' (scored as 1 through 7). Questions 2.11 and 2.14 were not used in the version of Questionnaire 2 that was given to astrophysicists because there is not a visible public policy debate concerning cosmic ray physics. Questions 2.15 and 2.16 were presented only to educated non-experts. Answer choices for these questions were 'None,' 'Very little,' 'Some,' and 'A great deal' (scored as 0 through 3). All participants were asked the same demographic questions as in Questionnaire 1.

We hypothesized that, on the basis of scientists' experience with ambiguous data, interpretive flexibility, and professional disagreements within their disciplines, scientists would give higher estimates than non-experts of how reasonable it can be for scientists to draw different conclusions from the same data set (Q2.2). We also tentatively hypothesized that scientists would be more circumspect or self-reflective when confronted with an equally informed but divergent opinion than non-experts (Q2.6 & Q2.7).

In the domain of morality, a number of philosophers (e.g., Ayer, 1936, Stevenson 1944) have argued that there are no correct answers to questions about what is morally right or wrong on the grounds that there is no objective way to resolve disagreements that arise concerning them. We hypothesized that persistent disagreement in climate science would lead lay people to have similar doubts about whether there are objective facts in this domain, or, at a minimum, whether they doubt that current scientific methods are capable of giving us knowledge of those facts (Q2.3, Q2.4 & Q2.5).

We hypothesized that educated non-experts would suspect that there was a greater degree of inappropriate silencing within academic and scientific institutions than scientists would (Q2.8 & Q2.9). We hypothesized that climate scientists would be more likely than non-experts to respond that the disagreements surrounding climate science did not actually fall within the science itself but rather lay in political debates about what to do in light of the science (Q2.11). We hypothesized that both experts and lay people would think that public discussions about the extent of disagreement within climate science have had a negative effect on public trust in science (Q2.13). Relatedly, we hypothesized that non-experts would express greater confidence in the results, methods, and authority of scientists working in other areas than of those working in climate science (Q2.15 & Q2.16).

## 3. Participants

We recruited a total of 3,367 participants for our study from January to April, 2017. Participants were invited to complete either Questionnaire 1 or Questionnaire 2, which were made available online. Climate scientists, climate policy experts, and astrophysicists were recruited via professional listservs and newsletters. Undergraduates at University College Dublin and alumni from the State University of New York at Buffalo were invited via email to participate in our study.[9] 23% of climate scientists reported having expertise in atmospheric science (e.g., meteorology, atmospheric physics), 44% in one of the earth sciences (incl. oceanography, glaciology, geology, hydrology), 44% in biological science (primarily ecology), and 5% in other sciences.[10] Climate policy experts had expertise in economics, law, political science, anthropology, resource management, conservation, agriculture, and philosophy.

---

[9] Additional information about participant recruitment and demographics can be found in the online supplementary materials document that accompanies this article.

[10] These percentages sum to more than 100% because several scientists indicated expertise in more than one of our areas of classification.

# 4. Results

## 4.1 Summary

The mean responses of the five participant groups to each of the items in Questionnaires 1 and 2 are summarized below in Tables 3 and 4.[11] For ease of reference, the rightmost column indicates the midpoint of the response scale for each item.

| Question | Climate scientists | Climate policy experts | Under-graduates | Alumni | Astro-physicists | Midpoint |
|---|---|---|---|---|---|---|
| **Q1.1** | 1.2 | 1.4 | 1.7 | 1.6 | 2.0 | 1.5 |
| **Q1.2** | 2.3 | 2.4 | 3.0 | 3.0 | 1.8 | 2 |
| **Q1.3** | 2.6 | 2.9 | 2.9 | 3.0 | 2.9 | 2 |
| **Q1.4** | 2.6 | 2.3 | 3.0 | 3.0 | 2.5 | 2 |
| **Q1.5** | 2.0 | 2.0 | 2.4 | 2.5 | 2.0 | 2 |
| **Q1.6** | 2.4 | 2.4 | 2.1 | 2.6 | 2.9 | 2 |
| **Q1.7** | 2.3 | 2.7 | 2.3 | 2.3 | 2.6 | 2 |
| **Q1.8** | 2.1 | 2.3 | 2.3 | 2.2 | 3.0 | 2 |
| **Q1.9** | 2.5 | 2.7 | 3.1 | 3.3 | 0.9 | 2 |
| **Q1.10** | 2.1 | 2.0 | 3.2 | 3.2 | 1.2 | 2 |
| **Q1.11** | 1.9 | 2.2 | 2.6 | 2.6 | 2.3 | 2 |
| **Q1.12** | 1.7 | 1.7 | 1.8 | 2.0 | 1.9 | 2 |
| **Q1.13** | 1.7 | 1.5 | 2.0 | 2.1 | 1.4 | 2 |

*Table 3*. Mean responses of the five participant groups to each of the items in Questionnaire 1, along with an additional column indicating the midpoint of the relevant response scale.

---

[11] Additional information about these statistics can be found in the supplementary materials document.

| Question | Climate scientists | Climate policy experts | Under-graduates | Alumni | Astro-physicists | Midpoint |
|---|---|---|---|---|---|---|
| Q2.1 | 1.0 | 1.5 | 1.8 | 1.6 | 2.0 | 1.5 |
| Q2.2 | 3.6 | 4.1 | 4.6 | 4.6 | 4.8 | 4 |
| Q2.3 | 3.1 | 3.5 | 4.0 | 4.0 | 4.8 | 4 |
| Q2.4 | 2.3 | 2.7 | 2.9 | 3.2 | 4.3 | 4 |
| Q2.5 | 3.3 | 3.7 | 4.6 | 4.2 | 4.1 | 4 |
| Q2.6 | 3.6 | 3.2 | 3.5 | 2.9 | 4.5 | 4 |
| Q2.7 | 5.0 | 5.0 | 4.7 | 4.8 | 4.1 | 4 |
| Q2.8 | 3.1 | 3.7 | 4.4 | 3.8 | 4.0 | 4 |
| Q2.9 | 2.7 | 3.4 | 4.2 | 3.8 | 3.2 | 4 |
| Q2.10 | 4.5 | 4.5 | 4.4 | 4.8 | 5.5 | 4 |
| Q2.11 | 5.1 | 5.2 | 4.8 | 4.8 | n/a | 4 |
| Q2.12 | 4.6 | 4.2 | 4.8 | 4.1 | 3.6 | 4 |
| Q2.13 | 3.6 | 3.1 | 3.4 | 3.6 | 3.7 | 4 |
| Q2.14 | 6.0 | 5.8 | 5.2 | 4.8 | n/a | 4 |
| Q2.15 | n/a | n/a | 2.6 | 2.5 | n/a | 1.5 |
| Q2.16 | n/a | n/a | 2.7 | 2.7 | n/a | 1.5 |

*Table 4*. Mean responses of the five participant groups to each of the items in Questionnaire 2, along with an additional column indicating the midpoint of the relevant response scale.

## 4.2 Disagreement

Questions 1.1 and 2.1 asked participants about the extent to which there is disagreement among experts about climate change or cosmic ray physics. Significant between-group differences were observed (cf. Figure 1).[12] Post-hoc tests performed on the answers of different pairs of participant groups indicated significant differences in every case, with the largest differences occurring between the answers of climate scientists and undergraduates ($r = .37$), climate scientists and alumni ($r = .23$), and climate scientists and astrophysicists ($r = .44$).[13] The mean response of climate scientists fell significantly below the midpoint of 1.5 ($r = .46$), whereas those of undergraduates ($r = .28$), alumni ($r = .11$), and astrophysicists ($r = .66$) fell significantly above the midpoint. The mean response of climate policy experts did not differ significantly from the midpoint.

---
[12] Kruskal-Wallis tests were used on each question to examine overall between-group differences. Pairwise differences were then examined with post-hoc Mann-Whitney tests. Cf. the supplementary materials document for further details concerning these tests.

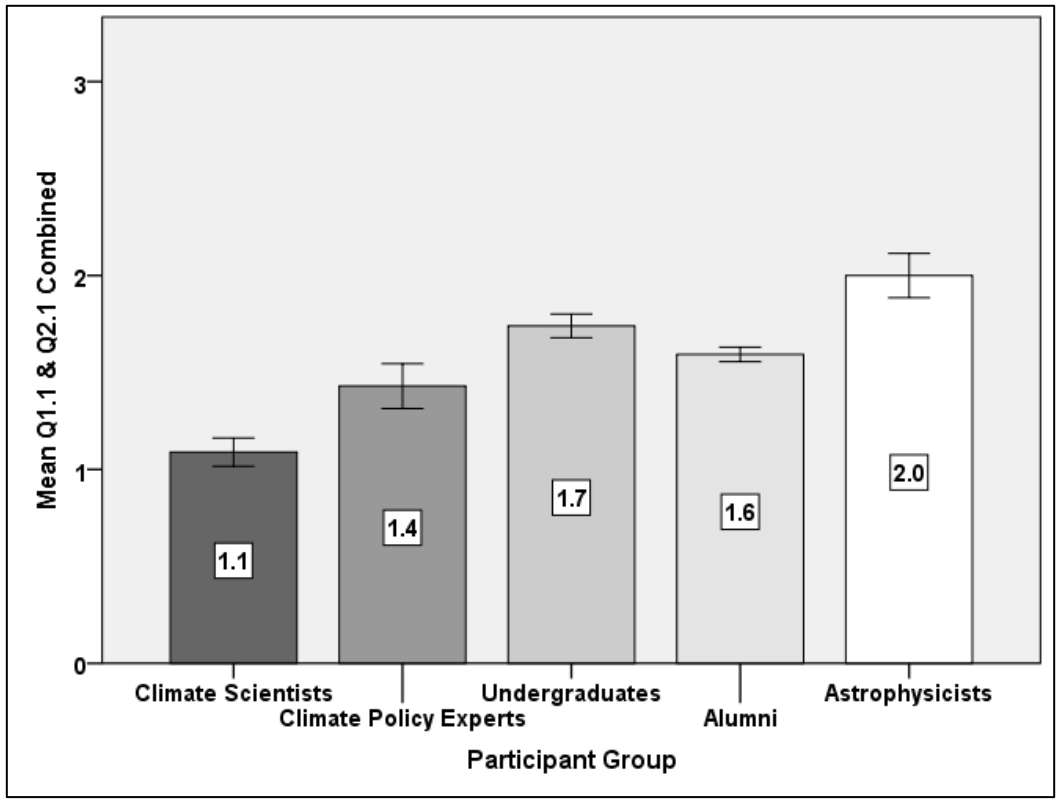

*Figure 1*. Mean responses to Questions 1.1 and 2.1 (combined), organized by participant group. Error bars in all figures represent 95% confidence intervals.

Many climate scientists indicated expertise in more than one of the following three broad categories: atmospheric science, earth science, and ecology. Focusing only on the 79% of climate scientists whose expertise fell squarely within only one of these categories, we observed significant between-group differences in how participants responded to Questions 1.1 and 2.1 (cf. Figure 2).[14] Post-hoc pairwise tests found significant differences between the responses of atmospheric scientists and ecologists ($r = .25$) and between earth scientists and ecologists ($r = .17$). Thus, within our sample of climate scientists, the closer a scientist's area of expertise was to what is considered to be the core of climate science, the greater the level of disagreement that scientist was likely to report there being within that field.

---

[13] Although perhaps most well-known as a measure of correlation, $r$ can be calculated as a measure of effect size for a variety of statistical tests.

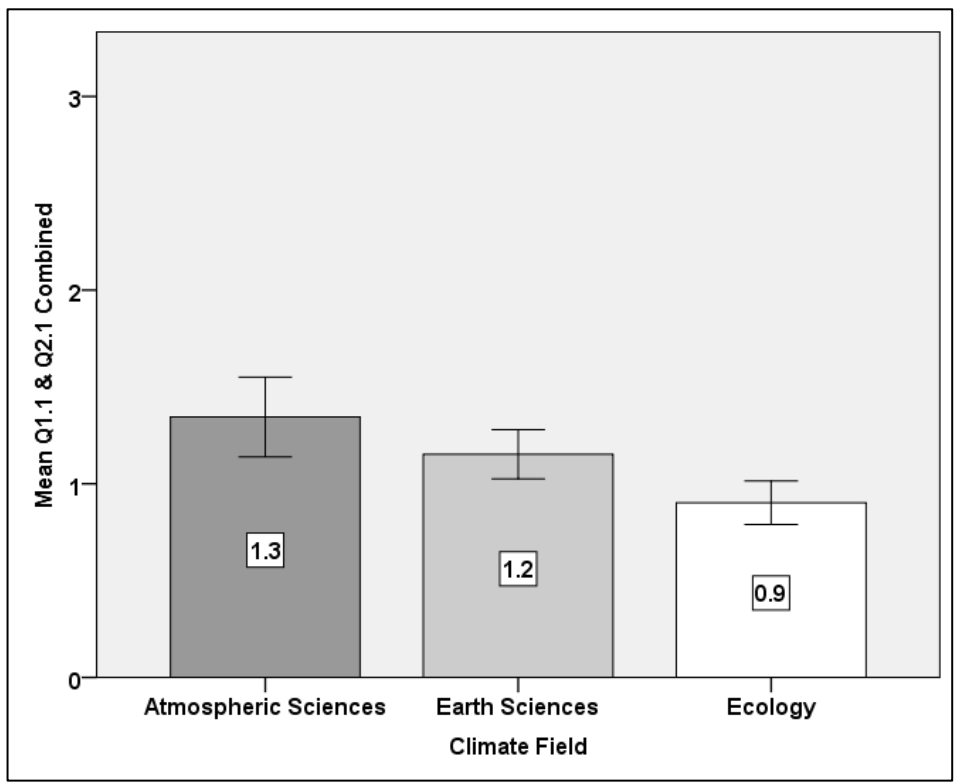

*Figure 2*. Mean responses to Questions 1.1 and 2.1 (combined) of those climate scientists whose areas of expertise falls into a single broad category.

---

[14] Kruskal-Wallis: $H(2) = 17.48$, $p < .001$. One-sample *t*-tests reveal that the mean response of atmospheric scientists did not differ significantly from the neutral midpoint of 1.5, whereas those of earth scientists ($r = .42$) and ecologists ($r = .64$) fell significantly below it.

One possible explanation of this interesting finding is that different aspects of the climate debate are salient to scientists from different disciplines and that these differences shaped their responses to our question. For example, a marine biologist who knows that rising global temperatures bring stress to the ecosystems she studies may focus on the fact that climate scientists do not disagree at all about whether the planet is warming. However, an atmospheric physicist participating in our study who works on 'climate sensitivity'—i.e., the question of how much global surface temperatures would rise given a doubling of atmospheric carbon dioxide—may respond to our question in light of the fact that different models of climate sensitivity yield different estimates.

It is widely noted that the debate about climate change in the United States is markedly different than in other parts of the world. In order to see whether our data reflected this difference, we sorted our participants (excluding astrophysicists) into the categories of expert (if they were climate scientists or climate policy experts) vs. non-expert and American vs. non-American. A two-way ANOVA revealed significant main effects for being an expert and being American and a significant interaction between the two variables.[15] In other words, being an American was a significant predictor of how participants answered Questions 1.1 and 2.1, but its predictive strength was greatest among the experts in our study.

---

[15] Expertise: $F(1, 3225) = 172.68$, $p < .00001$. Being American: $F(1, 3225) = 47.21$, $p < .00001$. Expertise * being American: $F(1, 3225) = 9.21$, $p < .01$.

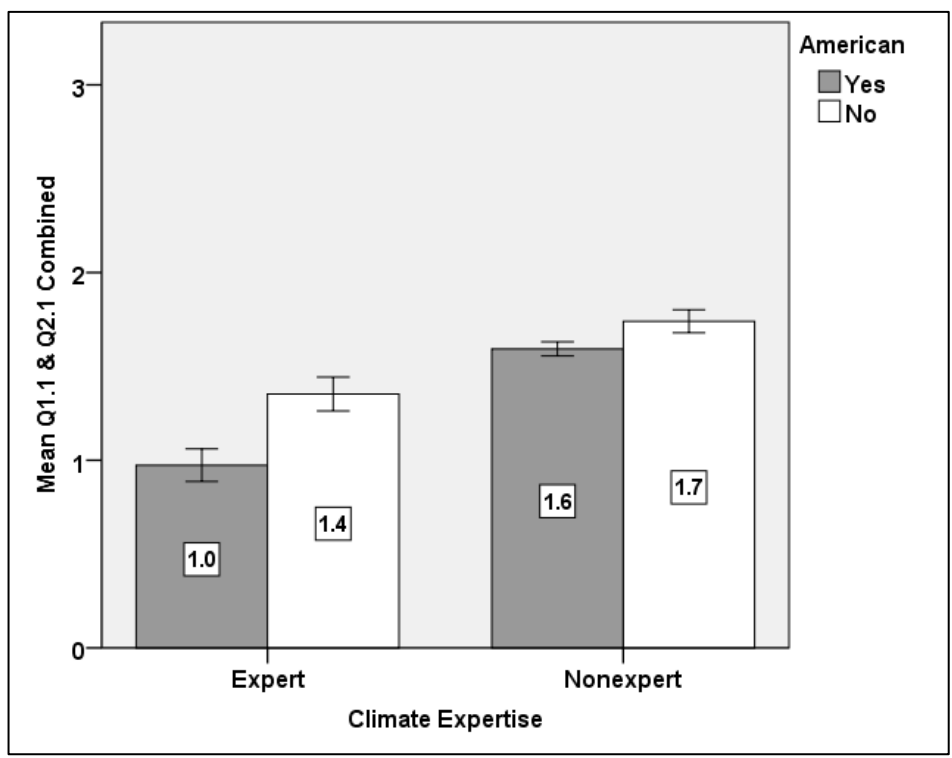

*Figure 3*. Mean responses to Questions 1.1 and 2.1 (combined), organized by whether or not participants are experts or American.

Participants' responses to Questions 1.1 and 2.1 may be shaped by the particular ways in which alleged disagreements within climate science are often leveraged for political ends, especially within the American context. Since experts in the U.S. know that any indication of disagreement or uncertainty about climate change that might appear in their responses could be used or even distorted by climate skeptics, they may feel pressure to downplay the extent of disagreement that they know exists. Another possibility is that different aspects of the climate debate may be salient to participants from different countries. In the U.S., admitting the very basic fact that the planet is warming is viewed as a strongly liberal stance. In other contexts where everyone agrees that the planet is warming, a question about the extent of disagreement within climate science may bring to mind other issues or questions, some of which may involve more uncertainty or disagreement than the basic question of increasing average temperature.

As compared to educated non-experts, climate experts more strongly agreed on Question 2.11 that disagreement within climate science concerned the public policy recommendations that should be made in light of the science rather than the science itself. However, the differences were not as large as the results from Questions 1.1 and 2.1 led us to expect. A two-way ANOVA performed on the responses of all participants except astrophysicists revealed small but statistically significant main effects for the variables of being a climate expert and being an American, and there was a significant interaction between being an expert and being an American.[16] American climate experts ($M = 5.4$) more strongly agreed that disagreements about climate change were most often about public policy questions than about core scientific questions than American non-experts ($M = 4.8$), non-American climate experts ($M = 4.9$), or non-American non-experts ($M = 4.8$).

**4.3 Epistemic Factors**

Questions 1.2 through 1.8 asked participants how important they thought various epistemic factors, as defined above, were to the emergence and persistence of disagreement about climate change or cosmic ray physics. The first half of these questions concerned the behaviors or attitudes of individual experts who are involved in scientific disagreements. The remaining questions in this section concerned the phenomena studied by scientific experts—e.g., their complexity, uncertainty in the data, and how difficult it is to obtain the right kind of data. On the whole, climate scientists and climate policy experts rated the epistemic factors as being less important contributors to expert disagreement than non-experts did.

---

[16] Expertise: $F(1, 1643) = 11.74$, $p < .001$. Being American: $F(1, 1643) = 5.27$, $p = .022$. Expertise * being American: $F(1, 1643) = 8.43$, $p < .01$.

Participants' ratings of the importance of the epistemic factors represented in Questions 1.2 through 1.5 positively correlated with one another to a noteworthy extent ($r$'s .23 to .46, all $p$'s < .000001), and their ratings of the factors in 1.6 through 1.8 correlated even more strongly ($r$'s .37 to .64, all $p$'s < .000001). Correlations between factors from the two groups correlated less strongly or failed to correlate significantly at all. Participants' ratings on Questions 1.4 through 1.8 also significantly predicted how much disagreement they thought there was within climate science or cosmic ray physics ($r$'s .18 to .36, all $p$'s < .000001).

There were no significant differences between the responses of the two groups of climate experts on any of these questions and the two groups of non-experts were generally in agreement about them. Importantly, the responses of climate experts in general differed from those of non-experts.

Post-hoc comparisons between climate scientists and astrophysicists revealed significant differences in their responses to Questions 1.2, 1.6, and 1.8 ($r$'s = .16, .15, .25), with astrophysicists being less likely to think that disagreements in their field were due to some of the experts being less well-informed and more likely to think that the issues about which experts disagree are very complex and that it is difficult to obtain enough of the right kind of data needed to resolve disagreements. In regard to the other epistemic factors, climate scientists and astrophysicists gave largely similar responses.

Experts in both fields indicated that there is agreement among experts about the methods, theoretical frameworks, and data that are needed to address the phenomena they study.

One reason the of climate experts to Question 1.2 fell significantly below those of educated non-experts may be that the general public is frequently exposed to poorly informed 'experts' through popular media—pundits who are in fact not climate scientists and thus not part of the community of scientists with which climate scientists would be having a scientific discussion. Hence, what gets perceived as expert disagreement by the public may not be taken as disagreement among genuine experts by those working within the field.

In response to Questions 1.3, 1.4, and 1.5, climate scientists and climate policy experts gave lower ratings of the extent to which expert disagreement within climate science was due to different starting points, focusing on different kinds of data, or differing views about the appropriate methods to use than did undergraduates or alumni.

Comparing the responses of experts and non-experts to Questions 1.2 through 1.5 to their responses to Questions 1.6 through 1.8, we see that both experts and non-experts largely agree that climate science involves uncertainties, deals with complex phenomena, and faces difficulties in obtaining the right kind of data. However, experts think that there is more theoretical or methodological agreement within their fields about how these epistemic fallibilities should be addressed or approached than non-experts do. Cosmic ray physicists gave largely similar answers to climate experts, except that the former rated the complexity of the phenomena they study and the difficulty of obtaining enough of the right kind of data about them as more important contributors to scientific disagreement.

**4.4 Nonepistemic Factors**

Items in the second half of Questionnaire 1 examined participants' judgments about the extent to which disagreements about climate change and cosmic ray physics are nurtured by political ideology (Q1.9), financial incentives (Q1.10), professional stake (Q1.11), careerism (Q1.12), or the psychological factor of stubbornness (Q1.13). The opinions of the non-experts paint a more negative picture of the state of climate science than those of practitioners within the field.

Participants' ratings of the importance of the nonepistemic factors represented in Questions 1.9 through 1.13 all strongly correlated with one another ($r$'s .30 to .70, all $p$'s < .000001). There were no significant differences between climate scientists and climate policy experts on any of the nonepistemic questions, and undergraduates and alumni differed slightly on only one question. By contrast, the responses of climate experts differed significantly from educated non-experts on almost all questions ($r$'s .09 to .38). The responses of astrophysicists differed significantly from those of all groups on Questions 1.9 and 1.10 ($r$'s .24 to .54) but exhibited fewer differences on the other questions.

The responses of scientific experts to Question 1.9 and 1.10 differed sharply from those of the educated non-experts, with the latter giving substantially higher ratings of the importance of these factors. All participant groups largely agreed that trying to make a name for oneself (Q1.12) and stubbornness, closed-mindedness, and unreasonableness (Q1.13) were considerably less significant nonepistemic factors.

For each participant, we averaged together their response to each question about epistemic factors (Q1.2 – Q1.8) and their response to each nonepistemic question (Q1.9 –

Q1.13), resulting in a composite epistemic score and a composite nonepistemic score. Mean composite scores for each group are represented in Figure 4. A two-way mixed ANOVA revealed significant main effects for participant group and composite score type (epistemic vs. nonepistemic), along with a significant interaction between group and score type.[17] In other words, (i) for a given composite score type, the scores of the groups differed significantly, (ii) for a given participant group, there were significant differences between their epistemic and nonepistemic scores, and (iii) the extent to which a participant group's epistemic and nonepistemic scores differed varied across groups.

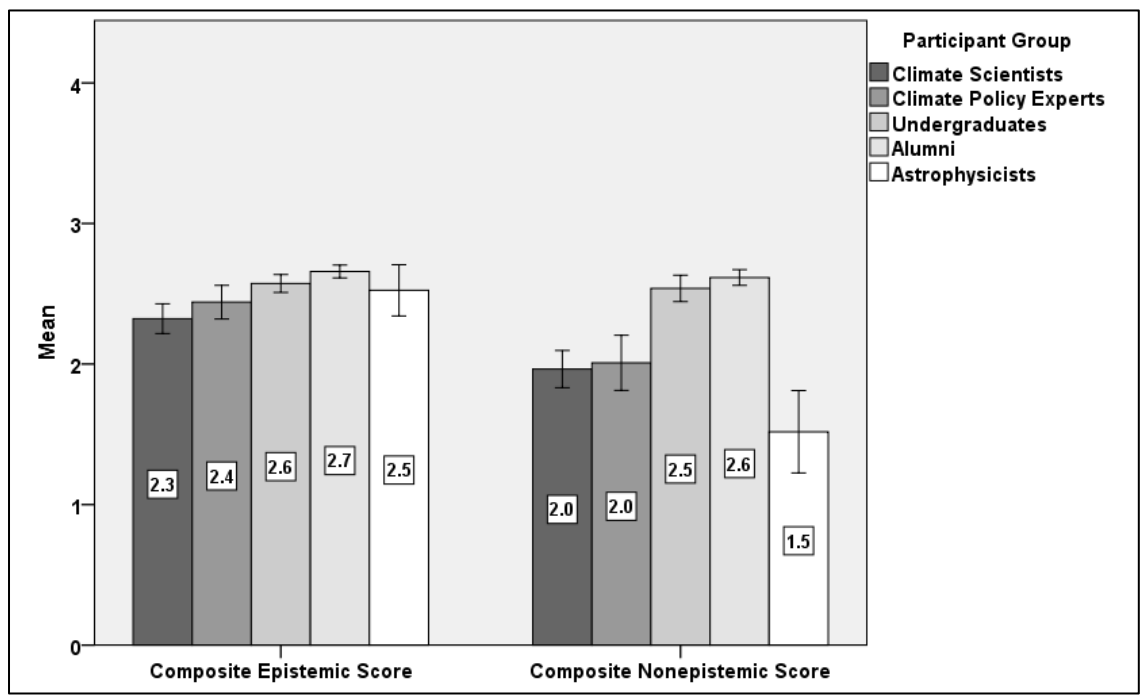

*Figure 4.* Mean composite scores for questions about epistemic and nonepistemic factors on Questionnaire 1, organized by participant group.

---

[17] Group: $F(4, 1624) = 38.59$, $p < .00001$. Score type: $F(1, 1624) = 83.14$, $p < .00001$. Group * score type: $F(4, 1624) = 15.90$, $p < .00001$.

The epistemic and nonepistemic composite scores were nearly identical for both undergraduates and alumni. In contrast, each of the three groups of experts gave lower ratings of the importance of nonepistemic. Notably, climate experts' estimates of the importance of nonepistemic factors were a good bit higher than those of astrophysicists.

We grouped all participants except astrophysicists into the categories of climate expert vs. non-expert and American vs. non-American and performed two two-way ANOVAs on participants' composite epistemic scores and their composite nonepistemic scores. With participants' composite epistemic scores, there was a significant main effect for being a climate expert and a significant interaction between being an expert and being an American.[18] The latter result was due to the fact that the largest pairwise differences on composite epistemic scores were observed between American experts and American non-experts. On participants' composite nonepistemic scores, there was only a significant main effect for being an expert.[19] On the whole, then, non-experts gave significantly higher ratings of the importance of both epistemic and nonepistemic factors in creating expert disagreement, and large differences were observed between experts and non-experts concerning non-epistemic factors such as political ideology, financial incentives, biases, and the need to defend one's life's work.

We believe that educated non-experts' ratings of the importance of nonepistemic factors indicates a significant lack of trust in professional scientists. In many studies of public trust in science (e.g., Pew 2015), members of the general public often indicate a lack of trust in particular scientific findings or in the ability of scientists to understand certain phenomena (e.g., health). One can get the impression that it is perhaps the scientific method that they distrust. However, at least in our data, the lack of trust that educated non-experts display is more strongly directed more toward individuals working within scientific institutions (as represented by participants' composite nonepistemic scores) than toward the methods employed by those scientists (as represented by their composite epistemic scores).

---

[18] Expert: $F(1, 1573) = 33.02$, $p < .000001$. Expert * American: $F(1, 1573) = 5.77$, $p = .016$.
[19] $F(1, 1573) = 87.18$, $p < .000001$.

**4.5 Normative Issues I: The Epistemology of Disagreement**

We turn now to data from Questionnaire 2 that concern normative epistemological issues such as whether it can be reasonable to draw different conclusions from the same data (Q2.2), whether participants decrease their confidence when faced with an equally well-informed person who disagrees with them (Q2.6), and whether participants conclude that someone with an atypical view is not actually as well informed as those with more standard views (Q2.7). Participants' responses to Questions 2.2, 2.6, and 2.7 were significantly correlated. Participants who took a more permissive stance toward evidence (Q2.2)—by indicating that two equally well-informed experts might look at the same data but reasonably arrive at different conclusions—were significantly more likely to take a conciliationist stance toward peer disagreement (Q2.6) ($r = .18$, $p < .000001$) and significantly less likely to judge that someone with a divergent opinion was not actually well-informed (Q2.7) ($r = -.20$, $p < .000001$). Participants who adopted a conciliationist attitude (Q2.6) were significantly less likely to judge that people holding divergent opinions were not well-informed (Q2.7) ($r = -.06$, $p = .02$).

On Question 2.2, the responses of climate scientists differed significantly from those of every other group ($r$'s .14 to .26), with climate scientists being more likely to deny there is more than one uniquely reasonable conclusion that can be drawn from a set of data. The responses of undergraduates, alumni, and astrophysicists did not differ significantly among themselves. Note that climate scientists were asked only about the science of climate change, so that it is possible that they might well adopt a more permissive approach to scientific evidence in other, less politicized domains.

On Question 2.6, the mean responses of climate scientists, climate policy experts, undergraduates, and alumni all fell significantly below the midpoint, indicating a disinclination to modify their views in light of contrary opinions. By contrast, astrophysicists were significantly inclined to modify their views upon learning about other experts who disagreed with them. Again, these data do not indicate that climate scientists and non-experts endorse a steadfast approach to peer disagreement across the board, since they might well adopt a more conciliationist approach in less politicized areas of science or everyday life. On Question 2.7, the mean responses of all four participant groups who were asked about climate change fell significantly above the midpoint, indicating a significant tendency to think that someone out of step with the mainstream was not really a well-informed expert after all. The mean response of astrophysicists did not differ from the midpoint.

**4.6 Normative Issues II: Objectivity**

Questions 2.3 through 2.5 probed participants' judgments about the ways in which expert disagreement might underwrite the conclusion that there is a lack of epistemological or metaphysical objectivity within the disputed domain. Participant groups were significantly divided on these questions. All pairwise post-hoc comparisons on Question 2.3 were significant ($r$'s .09 to .34), except for that between undergraduates and alumni. The mean responses of climate scientists, climate policy experts, undergraduates, and alumni all fell significantly below the midpoint on Question 2.4 ($r$'s .36 to .74), but there were important differences regarding how far below the midpoint these scores fell. The mean response of climate scientists to this question was their lowest on Questionnaire 2. In response to Question 2.5, the mean responses of undergraduates and alumni fell significantly above the midpoint, while that of climate scientists fell significantly above. Thus, climate scientists and climate policy experts strongly rejected the idea that the often exaggerated and politicized disagreements that exist should lead people to skeptical conclusions about whether there are objective facts about the nature of climate change, whether those facts can be known, and whether current scientific methods are equal to the task of uncovering them.

**4.7 Institutional Issues**

Questions 2.8, 2.9, and 2.10 examined participants' judgments about institutional biases and the benefits of expert disagreement. The responses of climate scientists differed significantly from those of every other group ($r$'s .16 to .79) on Questions 2.8 and 2.9. That is, climate experts indicated that they did not think the professional debate about climate change, as represented in academic journals and other venues, involved significant bias against minority views, with climate scientists expressing this opinion more strongly than climate policy experts. Undergraduates suspected to a certain degree that institutional bias was present, and alumni seemed uncertain about whether or not this was the case.

When we put the data from Questions 2.8 and 2.9 together with the data from the questions about nonepistemic factors in Section 4.4 above, we get the unfortunate conclusion that educated non-experts think there is more personal and institutional bias operative within climate science than do climate experts themselves.[20] These data seem to point to a crisis of trust in the men and women who work in climate science today.

The mean responses of all participant groups to Question 2.10 fell significantly above the midpoint, with that of astrophysicists being the highest. Thus, despite the fact that disagreements within science can sometimes be exploited for ideological aims, climate scientists and climate policy experts still agree that disagreement can be good for the health of their field.

There were significant correlations between participants' answers to Question 2.10 and their answers to 2.2 and 2.6. The more someone agreed that two experts could look at the same data but reasonably arrive at different conclusions (Q2.2), the more likely they were to think that disagreement could be good for a scientific discipline (Q2.10) ($r = .31$, $p < .000001$). Participants who reported taking a conciliationist approach to peer disagreement (Q2.6) were also significantly more likely to think disagreement could be good for science (Q2.10) ($r = .23$, $p < .000001$). These three questions thus all seem to tap into a kind of epistemic flexibility among our participants.

### 4.8 Public Trust in Science

As compared to other participant groups, climate scientists and undergraduates more strongly agreed that publicizing disagreement among scientific experts reduces public trust in science (Q2.12). The two groups of non-experts in our study were asked how much trust they have in climate science experts (Q2.15) and how much trust they have in scientists working in other areas of science (Q2.16). The mean response of each group fell significantly and substantially above the neutral midpoint for both questions ($r$'s .81 to .92—the largest effect sizes reported in this study). The confidence that each group expressed in climate scientists was significantly lower than the confidence they expressed in other scientists. There were no significant differences between the two participant groups. Thus, while non-experts express less confidence in climate scientists than other scientists, they express very high levels of confidence in both.

---

[20] We suspect that among non-experts without a college education this disparity may be even more pronounced.

**4.9 Public Policy**

The remaining two questions from Questionnaire 2 concerned the public policy implications of expert disagreement. The mean responses of all participants except astrophysicists to Question 2.13 fell significantly below the midpoint. Thus, participants disagreed to a modest extent that scientists should keep out of public policy discussions. All participant groups (excluding astrophysicists) agreed on Question 2.14 that the science of climate change was settled enough for policy purposes, with climate experts agreeing most strongly.

**5. Conclusion**

The goal of our study was to shed light on the role and consequences of disagreement among scientific experts and its implications for policy decisions and public opinion. We observed that, as compared with educated non-experts, climate experts believe (i) that there is less disagreement within climate science about climate change, (ii) that more of the disagreement that does exist concerns public policy questions rather than the science itself, (iii) that methodological factors play less of a role in generating existing disagreement among experts about climate science, (iv) that fewer personal and institutional biases influence the nature and direction of climate science research, (v) that there is more agreement among scientists about which methods or theoretical perspectives should be used to examine and explain the relevant phenomena, (vi) that disagreements about climate change should not lead people to conclude that the scientific methods being employed today are unreliable or incapable of revealing the truth, and (vii) that climate science is more settled than ideological pundits would have us believe and settled enough to base public policy on it. We also observed that the uniquely American political context predicted participants' judgments in many of these domains.

Our study also reveals that, as compared to climate scientists, astrophysicists working in cosmic ray physics were generally more willing to acknowledge expert disagreement, more open to the idea that a set of data can have multiple valid interpretations, and generally less quick to dismiss someone articulating a non-standard view as non-expert, than climate scientists. This was commensurate with the greater inherent uncertainty and data lacunae in their field. Experts in both climate science and astrophysics indicated that there is strong agreement within their fields about the methods, theoretical frameworks, and data that are needed to address the phenomena they study.

In line with existing studies of the opinions of climate scientists (Oreskes 2004, Bray & von Storch 2008, Anderegg et al. 2010, Cook et al. 2013, Stenhouse et al. 2014,

Cook et al. 2016) and the general public (Nisbet & Myers 2007, Leiserowitz et al. 2010, Leiserowitz et al. 2012, Pew 2012, Leiserowitz et al. 2015, Pew 2016, Gallup 2017), our findings show that despite the existence of a significant consensus on the causes of climate change among climate scientists, the general public continues to believe that climate scientists disagree over the fundamental cause of global warming. Our study goes beyond previous studies in examining (i) differences between expert and non-expert populations in regard to the factors these different groups think underlie expert disagreement in climate science, (ii) the attitudes of experts and non-experts toward normative issues of expert disagreement, and (iii) and how experts from politicized and depoliticized areas of science compare on various metrics. It is hoped that these results will inform our understanding of the nature and broader social consequences of expert disagreement.

<: Pew Research Center. 2015. "Public and Scientists' Views on Science and Society."
Available at:
http://www.pewinternet.org/files/2015/01/PI_ScienceandSociety_Report_012915.pdf.

Pew Research Center. 2016. "The Politics of Climate." Available at: http://www.pewinternet.org/2016/10/04/the-politics-of-climate/.

Stenhouse, N., Maibach, E., Cobb, S., Ban, R., Bleistein, A., Croft, P., Bierly, E., Seitter, K., Rasmussen, G., and Leiserowitz, A. 2014. "Meteorologists' Views About Global Warming: A Survey of American Meteorological Society Professional Members." *Bulletin of the American Meteorological Society* 95: 1029–40.

Stevenson, C. L. 1944. *Ethics and Language*. New Haven: Yale University Press.

**Divergent Perspectives on Expert Disagreement: Preliminary Evidence from Climate Science, Climate Policy, Astrophysics, and Public Opinion**

James R. Beebe (University at Buffalo), Maria Baghramian (University College Dublin), Luke Drury (Dublin Institute for Advanced Studies), Finnur Dellsén (Inland Norway University of Applied Sciences)

**Supplementary Materials**

**Peer Disagreement**

According to the recent philosophical literature, at a rough approximation, two or more people are epistemic peers in a given domain of enquiry *E* if (1) they have equal knowledge, training, cognitive skills such as powers of reasoning, epistemic virtues such as rationality and impartiality, intelligence, and background information, relevant to *E*, (2) there is no substantial difference in their cognitive abilities and limitations (3) they are equally well positioned to consider the available evidence regarding *E*, and (4) they have considered the available evidence regarding *E* with equal care and attention. Epistemic peers disagree when they have opposing and incompatible beliefs regarding *E*: two people have a disagreement if one of them believes a certain proposition *P* and the second disbelieves *P* or believes not *P*.

In cases of peer disagreement, most contributors to the debate defend some version of the view that one should move closer to one's peers' opinion, e.g., by suspending judgment or by adopting an intermediate level of confidence between the disagreeing peer and one's former self (e.g., Christensen 2006, Feldman 2006, Elga 2007). This family of views is known as *conciliationism*. In contrast, *steadfastness* holds that one should 'stand one's ground' in the face of peer disagreement, i.e., continue to have the same beliefs and levels of confidence as one did before the disagreement. Although this is certainly a minority view in the literature, it does have its proponents (e.g., Kelly 2005, 2010).

Question 2.2 asks a general question about the possibility of reasonable and apparently faultless disagreement—i.e., disagreements where neither side seems to be making any obvious errors and disputants are equally well informed and possess the same evidence. It concerns the following principle that figures centrally in philosophical discussions of how one should respond to equally well informed peers with whom one disagrees:

> *Uniqueness*. In any given evidential situation, there is only one attitude that it is rational to take toward a proposition in light of the evidence one possesses.

Cf. Christensen (2009) for an overview of how a principle like this figures in contemporary philosophical debates, although Christensen focuses his attention on a slightly stronger version of Uniqueness.

**Participants**

Climate scientists were recruited via messages to the following listservs: ecolog-l (maintained by the Ecological Society of America), coral-list (maintained by the National Oceanic and Atmospheric Administration's Coral Reef Conservation Program and Atlantic Oceanographic and Meteorological Laboratory), cryolist (affiliated with the International Glaciological Society), arcticinfo (for arctic climate scientists), and marine-b (for marine biodiversity scientists). Climate policy experts were recruited via messages to climate-l (maintained by the International Institute for Sustainable Development), the primary listserv for climate policy experts. Astrophysicists were recruited via an invitation in a bulletin of the International Astronomical Union Division D (High Energy Phenomena and Fundamental Physics) and listservs for researchers affiliated or collaborating with scientists at the Dublin Institute for Advanced Study, the Pierre Auger Observatory (Argentina), and the High Energy Stereoscopic System (Namibia). Undergraduates were recruited via email with assistance from the UCD IT Services and Student Services offices. Alumni from the University at Buffalo were recruited via email with assistance from the UB Office of Development and Alumni Communications.

|  | Total | Ave. Age | % Female | Ave. Years Experience | % with Doctoral Degree | % with Master's Degree only |
|---|---|---|---|---|---|---|
| **Climate scientists** | 457 | 43 | 41 | 17 | 66 | 26 |
| **Climate policy experts** | 200 | 48 | 31 | 18 | 54 | 38 |
| **Undergraduates** | 697 | 23 | 52 | n/a | n/a | n/a |
| **Alumni** | 1,914 | 52 | 44 | n/a | 28 | 40% |
| **Astrophysicists** | 99[1] | 49 | 17 | 22 | 91 | 4 |

*Table S1*. Participant demographics.

|  | UK or Ireland | Rest of Europe | U.S. | Rest of Americas | Other |
|---|---|---|---|---|---|
| **Climate scientists** | 6% | 28% | 45% | 8% | 13% |
| **Climate policy experts** | 14% | 23% | 30% | 10% | 23% |
| **Astrophysicists** | 6% | 59% | 17% | 9% | 9% |

*Table S2*. Nationalities of expert participants.

---

[1] The total global community of cosmic ray physicists is only a few hundred, so this number represents a good proportion of the scientists in this field.

Because the various experts we recruited were from several different countries and because we suspected that differences in judgments might fall along national lines, we asked experts to indicate their nationality. Because the university undergraduates we recruited were all attending school in Ireland and all of the university alumni attended school in New York, we asked them about their ethnicity rather than about their nationalities. The undergraduates reported that they were 83% white Irish, and the alumni were 82% Caucasian.

**Additional Analyses, Figures, and Tables**

D'Agostino-Pearson omnibus tests for normality (D'Agostino 1986) were performed on the distributions of each participant group's answers to each question. The D'Agostino-Pearson test works better for large samples than the Kolmogorov-Smirnov or Shapiro-Wilk tests. The D'Agostino-Pearson tests, along with visual inspection of histograms and P-P plots, revealed that a large majority of the distributions of responses were non-normal. The distributions of responses from climate scientists, undergraduates, and alumni were non-normal on every question, and approximately half of the answer distributions of climate policy experts and astrophysicists were non-normal. Shapiro-Wilk tests for normality, which are better suited for smaller sample sizes, were also performed of the distributions of responses from climate policy experts and astrophysicists. These tests indicated that the relevant answer distributions were non-normal. Logarithmic and square root transformations of the data failed to result in normal distributions. Therefore, with only a few exceptions, non-parametric tests were used in the analyses below when comparing the answers of different groups to the same question or the answers of members of the same group to different questions. Parametric one-sample *t*-tests were used above to test whether the mean response of each of the five participant groups to particular items differed from the neutral midpoint because there is not a good non-parametric equivalent and because it is the normality of sampling distributions rather than of sample data that is most central to the validity of the *t*-test.

Kruskal-Wallis tests were used on each question to examine overall between-group differences. Pairwise differences were then examined with post-hoc Mann-Whitney tests, and Benjamini-Hochberg corrections were used in each case to control for multiple comparisons.

The mean responses of the five participant groups to each of the items in Questionnaires 1 and 2 are summarized below in Tables S3 and S4. For ease of reference, the rightmost column indicates the midpoint of the response scale for each item. One-sample *t*-tests were run on the set of responses represented by each cell, in order to see whether they differed significantly from the relevant midpoint. All statistical tests reported in this paper are two-tailed. To correct for multiple comparisons, initial p values were adjusted using the Benjamini-Hochberg method. Mean responses whose adjusted *p* values were less than .05 are marked with a '*', those that were less than .01 are marked with a '**', and those below .001 are marked with a '***'.

| Question | Climate scientists | | | Climate policy experts | | | Undergraduates | | | Alumni | | | Astrophysicists | | | Midpoint |
|---|---|---|---|---|---|---|---|---|---|---|---|---|---|---|---|---|
| | *M* | *SD* | *r* | *M* | *SD* | *r* | *M* | *SD* | *r* | *M* | *SD* | *r* | *M* | *SD* | *r* | |
| **Q1.1** | 1.2*** | 0.8 | .39 | 1.4 | 0.8 | n/a | 1.7*** | 0.8 | .27 | 1.6*** | 0.8 | .12 | 2.0*** | 0.5 | .69 | 1.5 |
| **Q1.2** | 2.3** | 1.3 | .22 | 2.4** | 1.2 | .33 | 3.0*** | 1.1 | .70 | 3.0*** | 1.1 | .68 | 1.8 | 1.3 | n/a | 2 |
| **Q1.3** | 2.6*** | 1.2 | .47 | 2.9*** | 1.0 | .69 | 2.9*** | 1.0 | .65 | 3.0*** | 1.0 | .72 | 2.9*** | 1.0 | .69 | 2 |
| **Q1.4** | 2.6*** | 1.2 | .43 | 2.3* | 1.2 | .26 | 3.0*** | 1.0 | .72 | 3.0*** | 1.0 | .71 | 2.5** | 1.0 | .42 | 2 |
| **Q1.5** | 2.0 | 1.3 | n/a | 2.0 | 1.2 | n/a | 2.4*** | 1.2 | .31 | 2.5*** | 1.2 | .38 | 2.0 | 1.3 | n/a | 2 |
| **Q1.6** | 2.4*** | 1.2 | .30 | 2.4*** | 1.1 | .37 | 2.1* | 1.2 | .12 | 2.6*** | 1.2 | .43 | 2.9*** | 1.1 | .64 | 2 |
| **Q1.7** | 2.3** | 1.3 | .20 | 2.7*** | 1.1 | .52 | 2.3*** | 1.2 | .24 | 2.3*** | 1.3 | .25 | 2.6*** | 1.0 | .53 | 2 |
| **Q1.8** | 2.1 | 1.4 | n/a | 2.3 | 1.2 | n/a | 2.3*** | 1.3 | .20 | 2.2*** | 1.4 | .14 | 3.0*** | 0.9 | .76 | 2 |
| **Q1.9** | 2.5*** | 1.5 | .29 | 2.7*** | 1.4 | .45 | 3.1*** | 1.2 | .66 | 3.3*** | 1.1 | .76 | 0.9*** | 1.2 | .66 | 2 |
| **Q1.10** | 2.1 | 1.5 | n/a | 2.0 | 1.5 | n/a | 3.2*** | 1.1 | .73 | 3.2*** | 1.2 | .71 | 1.2*** | 1.4 | .52 | 2 |
| **Q1.11** | 1.9 | 1.3 | n/a | 2.2 | 1.3 | n/a | 2.6*** | 1.2 | .45 | 2.6*** | 1.3 | .42 | 2.3 | 1.1 | n/a | 2 |
| **Q1.12** | 1.7*** | 1.2 | .25 | 1.7 | 1.3 | n/a | 1.8** | 1.3 | .14 | 2.0 | 1.3 | n/a | 1.9 | 1.2 | n/a | 2 |
| **Q1.13** | 1.7** | 1.3 | .23 | 1.5** | 1.3 | .35 | 2.0 | 1.4 | n/a | 2.1 | 1.3 | n/a | 1.4** | 1.2 | .48 | 2 |

*Table S3*. Mean responses of the five participant groups to each of the items in Questionnaire 1, along with standard deviations, a measure of effect size, and an additional column indicating the midpoint of the relevant response scale. *R* values represent the size of the mean's difference from the neutral midpoint.

S7| Question | Climate scientists | | | Climate policy experts | | | Undergraduates | | | Alumni | | | Astrophysicists | | | Midpoint |
|---|---|---|---|---|---|---|---|---|---|---|---|---|---|---|---|---|
| | *M* | *SD* | *r* | *M* | *SD* | *r* | *M* | *SD* | *r* | *M* | *SD* | *r* | *M* | *SD* | *r* | |
| Q2.1 | 1.0*** | 0.8 | .52 | 1.5 | 0.8 | n/a | 1.8*** | 0.8 | .30 | 1.6** | 0.8 | .11 | 2.0*** | 0.6 | .63 | 1.5 |
| Q2.2 | 3.6*** | 1.9 | .24 | 4.1 | 1.9 | n/a | 4.6*** | 1.7 | .32 | 4.6*** | 1.8 | .33 | 4.8*** | 1.5 | .45 | 4 |
| Q2.3 | 3.1*** | 1.8 | .46 | 3.5** | 1.9 | .27 | 4.0 | 1.8 | n/a | 4.0 | 1.9 | n/a | 4.8** | 1.7 | .40 | 4 |
| Q2.4 | 2.3*** | 1.6 | .74 | 2.7*** | 1.8 | .58 | 2.9*** | 1.8 | .52 | 3.2*** | 2.0 | .36 | 4.3 | 1.8 | n/a | 4 |
| Q2.5 | 3.3*** | 1.9 | .35 | 3.7 | 2.0 | n/a | 4.6*** | 1.8 | .34 | 4.2** | 1.9 | .10 | 4.1 | 1.9 | n/a | 4 |
| Q2.6 | 3.6*** | 1.6 | .27 | 3.2*** | 1.7 | .43 | 3.5*** | 1.7 | .27 | 2.9*** | 1.6 | .57 | 4.5* | 1.5 | .33 | 4 |
| Q2.7 | 5.0*** | 1.5 | .56 | 5.0*** | 1.6 | .53 | 4.7*** | 1.6 | .41 | 4.8*** | 1.7 | .45 | 4.1 | 1.6 | n/a | 4 |
| Q2.8 | 3.1*** | 1.7 | .46 | 3.7 | 1.6 | n/a | 4.4*** | 1.6 | .21 | 3.8** | 1.8 | .10 | 4.0 | 1.6 | n/a | 4 |
| Q2.9 | 2.7*** | 1.7 | .63 | 3.4** | 1.9 | .30 | 4.2 | 1.8 | n/a | 3.8*** | 1.8 | .12 | 3.2*** | 1.6 | .44 | 4 |
| Q2.10 | 4.5*** | 1.8 | .25 | 4.5** | 1.9 | .27 | 4.4*** | 1.9 | .20 | 4.8*** | 1.8 | .42 | 5.5*** | 1.5 | .70 | 4 |
| Q2.11 | 5.1*** | 1.5 | .59 | 5.2*** | 1.8 | .57 | 4.8*** | 1.5 | .48 | 4.8*** | 1.6 | .43 | n/a | n/a | n/a | 4 |
| Q2.12 | 4.6*** | 1.9 | .31 | 4.2 | 1.9 | n/a | 4.8*** | 1.8 | .39 | 4.1 | 1.8 | n/a | 3.6 | 1.7 | n/a | 4 |
| Q2.13 | 3.6** | 2.0 | .19 | 3.1*** | 1.9 | .45 | 3.4*** | 1.8 | .31 | 3.6*** | 2.0 | .20 | 3.7 | 1.9 | n/a | 4 |
| Q2.14 | 6.0*** | 1.7 | .76 | 5.8*** | 1.7 | .74 | 5.2*** | 1.6 | .60 | 4.8*** | 2.1 | .38 | n/a | n/a | n/a | 4 |
| Q2.15 | n/a | n/a | n/a | n/a | n/a | n/a | 2.6*** | 0.6 | .88 | 2.5*** | 0.7 | .81 | n/a | n/a | n/a | 1.5 |
| Q2.16 | n/a | n/a | n/a | n/a | n/a | n/a | 2.7*** | 0.5 | .92 | 2.7*** | 0.5 | .92 | n/a | n/a | n/a | 1.5 |

*Table S4*. Mean responses of the five participant groups to each of the items in Questionnaire 2, along with standard deviations, a measure of effect size, and an additional column indicating the midpoint of the relevant response scale. *R* values represent the size of the mean's difference from the neutral midpoint.

| Question | df | H | P |
|---|---|---|---|
| Q1.1 | 4 | 83.537 | < .000001 |
| Q1.2 | 4 | 99.78 | < .000001 |
| Q1.3 | 4 | 27.14 | < .0001 |
| Q1.4 | 4 | 61.76 | < .000001 |
| Q1.5 | 4 | 40.56 | < .000001 |
| Q1.6 | 4 | 40.09 | < .000001 |
| Q1.7 | 4 | 8.77 | .067 |
| Q1.8 | 4 | 16.83 | < .01 |
| Q1.9 | 4 | 146.64 | < .000001 |
| Q1.10 | 4 | 207.4 | < .000001 |
| Q1.11 | 4 | 54.89 | < .000001 |
| Q1.12 | 4 | 14.64 | < .01 |
| Q1.13 | 4 | 34.51 | < .000001 |

*Table S5*. Results of Kruskal-Wallis tests for participants' responses to Questionnaire 1.

| Question | df | H | P |
|---|---|---|---|
| Q2.1 | 4 | 130.64 | < .000001 |
| Q2.2 | 4 | 65.71 | < .000001 |
| Q2.3 | 4 | 72.42 | < .000001 |
| Q2.4 | 4 | 79.25 | < .000001 |
| Q2.5 | 4 | 72.81 | < .000001 |
| Q2.6 | 4 | 92.08 | < .000001 |
| Q2.7 | 4 | 18.95 | < .01 |
| Q2.8 | 4 | 72.97 | < .000001 |
| Q2.9 | 4 | 110.68 | < .000001 |
| Q2.10 | 4 | 30.64 | < .0001 |
| Q2.11 | 3 | 19.73 | < .001 |
| Q2.12 | 4 | 55.14 | < .000001 |
| Q2.13 | 4 | 9.31 | .054 |
| Q2.14 | 3 | 112.28 | < .000001 |

*Table S6*. Results of Kruskal-Wallis tests for participants' responses to Questionnaire 2.

As compared to educated non-experts, climate experts more strongly agreed on Question 2.11 that disagreement within climate science concerned the public policy recommendations that should be made in light of the science rather than the science itself. However, the differences were not as large as the results from Questions 1.1 and 2.1 led us to expect. Each pairwise comparison between one expert group and one non-expert group was significant, but the effect sizes were small ($r$'s between .09 and .14).

On Questions 1.2 through 1.8 about epistemic factors that contribute to expert disagreement, post-hoc Mann-Whitney comparisons revealed that the two groups of non-experts differed only on Questions 1.3 ($r = .07$, a statistically significant but theoretically negligible effect size) and 1.6 ($r = .16$). Pairwise comparisons also indicated that scientists differed significantly from undergraduates on all questions ($r$'s from .11 to .27), except 1.6, and from alumni on all questions ($r$'s from .15 to .20), except 1.6 and 1.8. The answers of climate policy experts differed significantly from undergraduates on all questions ($r$'s from .11 to .23), except 1.3 and 1.8, and from alumni on all questions ($r$'s from .12 to .17), except 1.3, 1.6, and 1.8.

Post hoc pairwise comparisons on Questions 1.9 through 1.13, which concerned nonepistemic factors that give rise to expert disagreement, revealed that the responses of climate scientists differed significantly from those of undergraduates on all questions ($r$'s .12 to .38), except 1.12, which was about careerism. Climate scientists differed significantly from alumni on all questions ($r$'s .09 to .30). The responses of climate policy experts differed significantly from those of both groups non-experts on all questions, except Question 1.12 ($r$'s .11 to .34). The responses of undergraduates and alumni to Questions 1.9 through 1.13 differed significantly on the role of political ideology ($r = .08$), but this theoretically negligible effect size achieved significance only because 1,290 participants were being compared in the analysis. The answers of undergraduates and alumni did not differ significantly on any of the other nonepistemic factors. Astrophysicists' responses did not differ significantly from any group on Questions 1.11 and 1.12. On Question 1.13, astrophysicists did not differ significantly from climate scientists and climate policy experts but did differ from undergraduates ($r = .15$) and alumni ($r = .11$) to a small extent. Undergraduates and alumni gave higher ratings to the importance of defending theories that represent one's life's work (Q1.11) than the three groups of experts.

The mean composite epistemic score of each participant group fell significantly above the midpoint ($r$'s .39 to .68). However, the mean composite nonepistemic scores of climate scientists and climate policy experts failed to differ from the midpoint, while those of undergraduates ($r = .51$) and alumni ($r = .58$) fell significantly above the midpoint, and those of astrophysicists ($r = .45$) fell significantly below.

|  |  | Mean Composite Epistemic Score | | Mean Composite Nonepistemic Score | |
|---|---|---|---|---|---|
|  |  | American | Non-American | American | Non-American |
| **Climate Expertise** | **Expert** | 2.3 | 2.4 | 2.1 | 2.0 |
|  | **Non-Expert** | 2.7 | 2.6 | 2.6 | 2.5 |

*Table S7.* Mean composite scores, grouped by epistemic vs. nonepistemic, expert vs. non-expert, and American vs. non-American.

A two-way (climate expert/non-expert x American/non-American) ANOVA performed on participants' responses to Question 2.2 (excluding astrophysicists) failed to be deliver a result of significance. A similar analysis performed on Question 2.6 revealed that American climate experts, non-American climate experts, and non-American non-experts all had identical mean scores (3.5), whereas American non-experts had a significantly lower mean score (2.9). This difference resulted in there being significant main effects for expertise and being American, and a significant interaction between the two. (Expertise: $F(1, 1641) = 7.76, p < .01$. Being American: $F(1, 1641) = 8.94, p < .01$. Expertise * being American: $F(1, 1641) = 9.51, p < .01$.) On Question 2.7, a two-way ANOVA revealed a significant main effect for expertise and a nearly significant effect for being an American. (Expertise: $F(1, 1640) = 4.53, p = .033$. Being American: $F(1, 1640) = 3.54, p = .06$.)

The mean response of astrophysicists to Question 2.4 fell near the midpoint. In response to Question 2.5, the mean responses of undergraduates ($r = .34$) and alumni ($r = .10$) fell significantly above the midpoint, while that of climate scientists fell significantly above ($r = .35$).

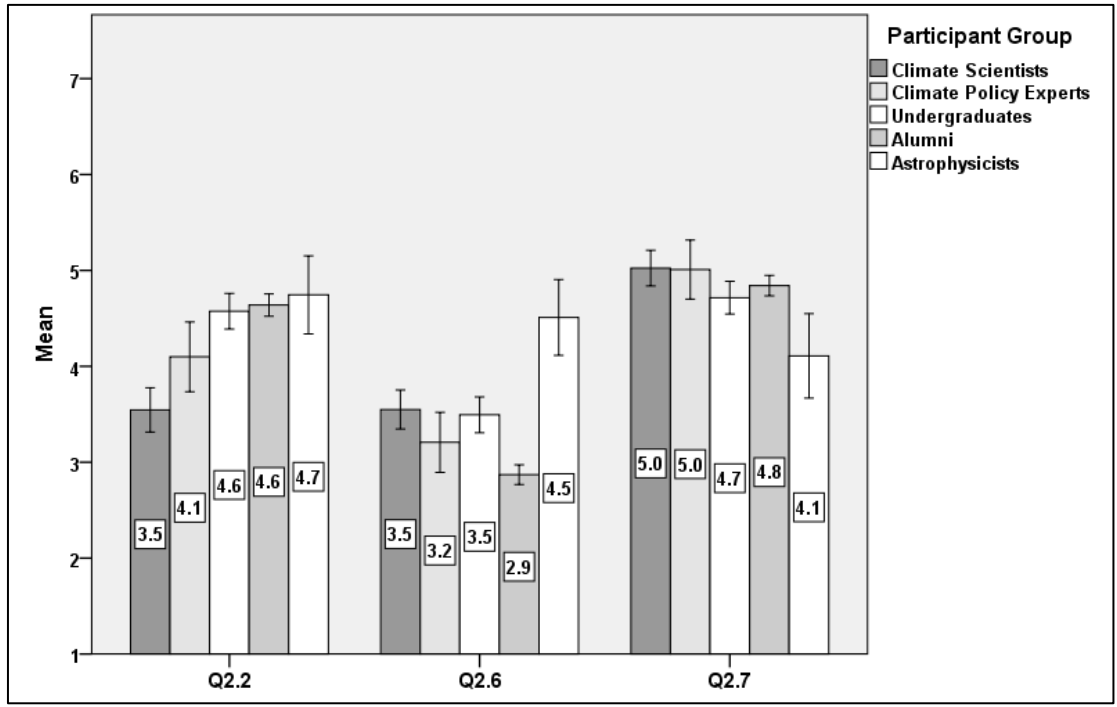

*Figure S1.* Mean responses to Questions 2.2, 2.6, and 2.7, organized by participant group.

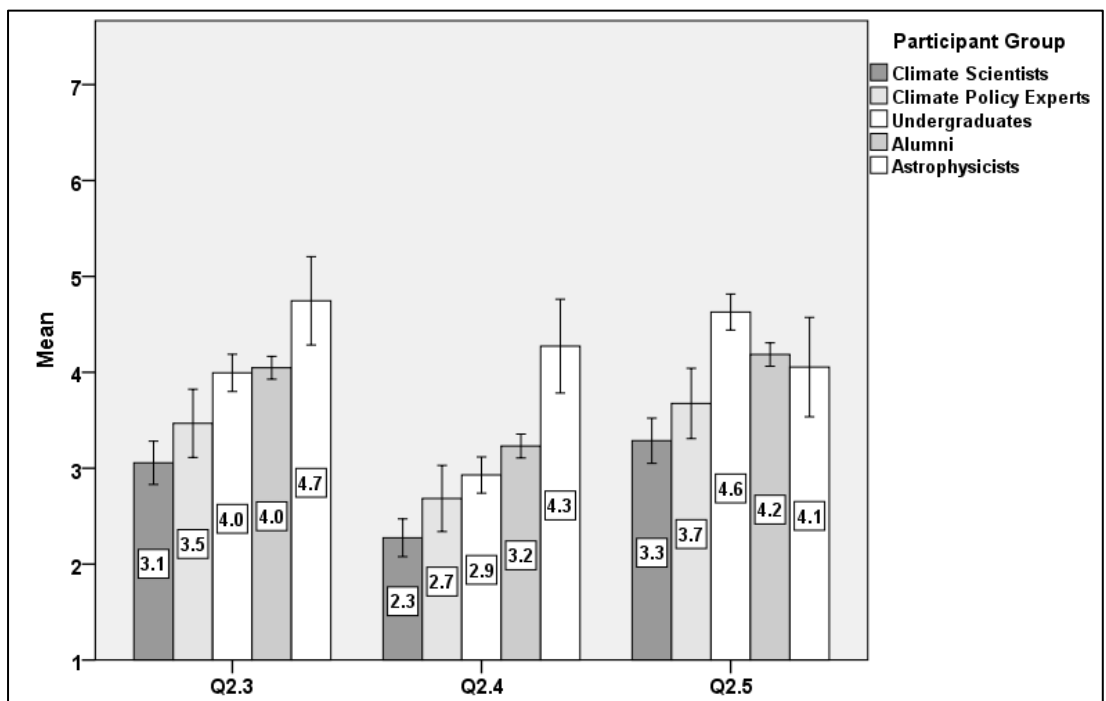

*Figure S3.* Mean responses to Questions 2.3, 2.4, and 2.5, organized by participant group.

S13

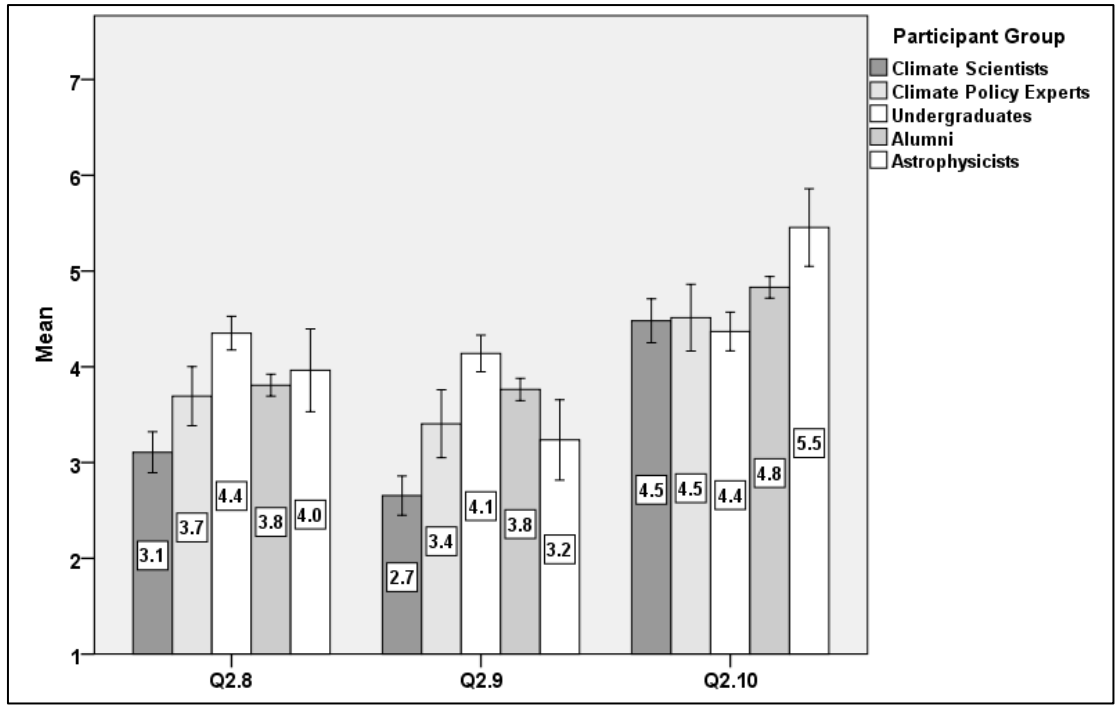

*Figure S4.* Mean responses to Questions 2.8, 2.9, and 2.10, organized by participant group.

We averaged participants' responses to Questions 2.8 and 2.9 to obtain an institutional bias score. We then combined all participants who were asked about climate science and sorting them by whether or not they were climate experts or American. A two-way ANOVA revealed significant main effects for being an expert and being an American, but the latter effect was in the opposite direction than what we had predicted. (Expertise: $F(1, 1642) = 92.186$, $p < .0001$. Being American: $F(1, 1642) = 16.71$, $p < .000001$.) Among both climate experts and educated non-experts, Americans were less likely to judge there was institutional bias against controversial hypotheses or minority views than non-Americans.

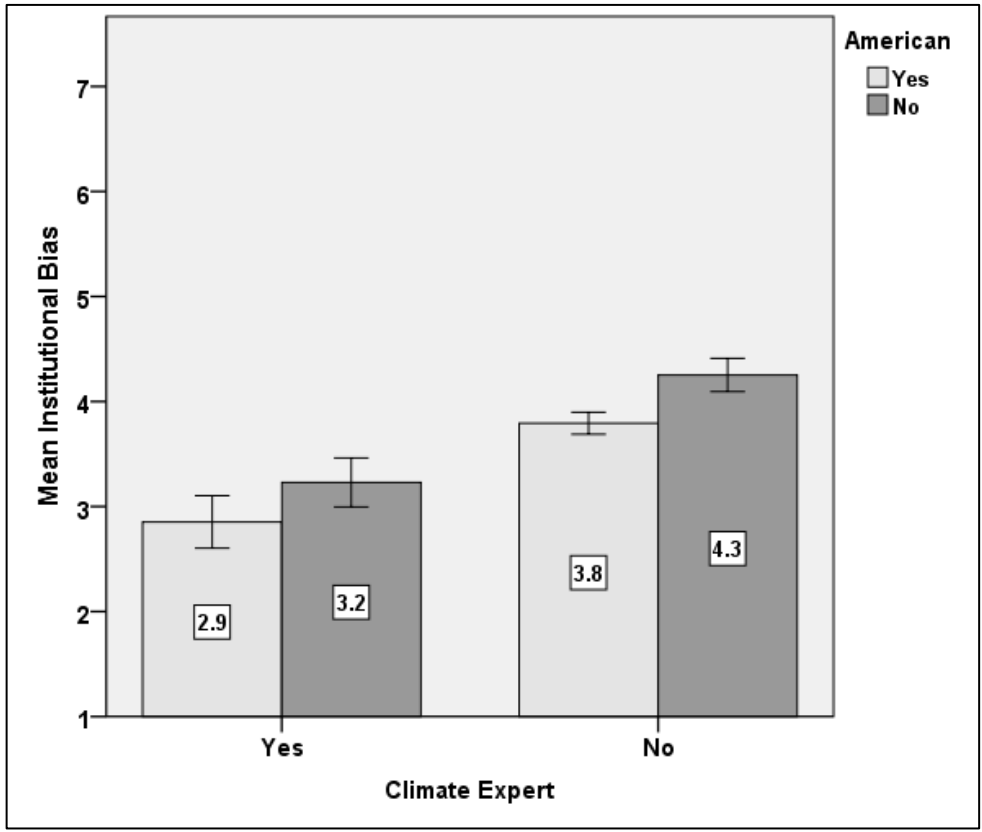

*Figure S2.* Mean institutional bias scores for all participants except astrophysicists, organized by whether or not participants are climate experts or American.

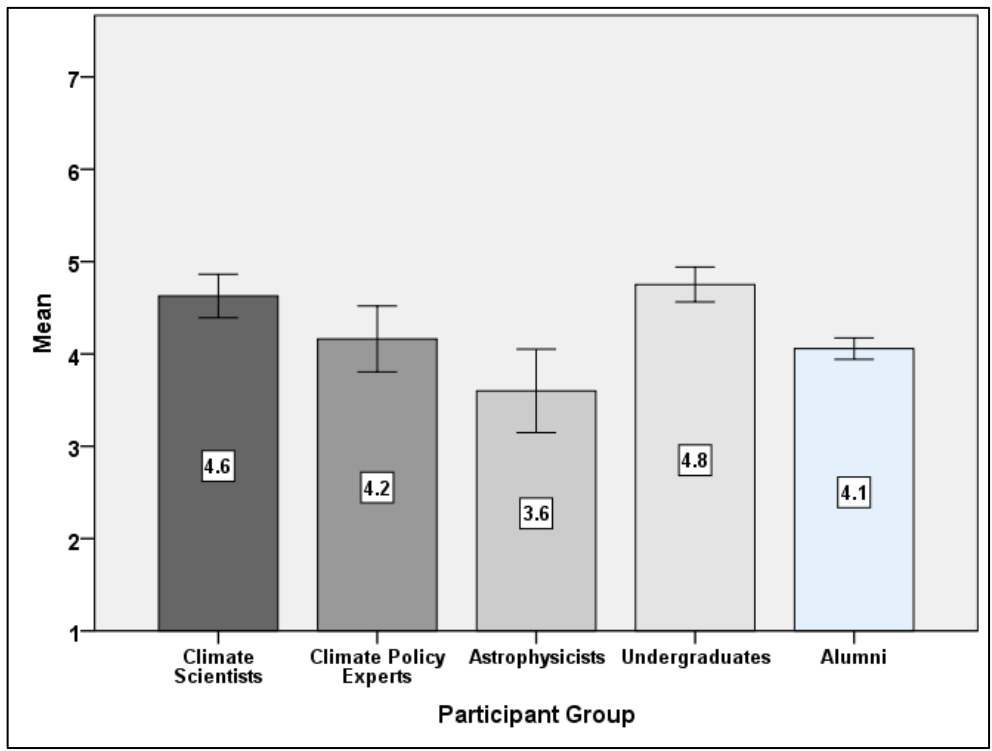

*Figure S5.* Mean participant responses to Question 2.12, organized by participant group.

On Question 2.14, the differences between the answers of the two groups of climate experts differed significantly from the answers of each of our groups of non-experts (*r*'s .17 to .30).

For Questions 2.15 and 2.16, Wilcoxon signed-rank tests were used to test for between-question differences, and Mann-Whitney tests were used to test for between-group differences. Both undergraduates ($r = -.22$) and alumni ($r = -.29$) showed significant between-question differences (Undergraduates: $z = -4.04, p < .0001$. Alumni: $z = -9.01, p < .00001$). The Benjamini-Hochberg method was used to correct for multiple comparisons.

**Disagreement in Cosmic Ray Physics**

The questions we asked about cosmic ray physics concerned the science as a whole and did not introduce finer distinctions between specific topics or areas of disagreement. Because this area of research is likely to be unfamiliar to many readers it may be useful to give a brief overview. The surprising existence of a highly penetrating ionising radiation reaching the Earth from outer space was discovered over one hundred years ago, but many features of these "cosmic rays" are still poorly understood.

Observationally the main issue in cosmic ray physics is that the flux of particles falls off extremely rapidly as one goes to higher particle energies so that no one experiment works over more than a small part of the energy range that needs to be covered. The flux of mildly relativistic particles above the atmosphere is about one per square centimetre per second so that quite small satellite experiments can be used, whereas at the very highest energies we see only about one particle per square kilometre per century requiring large experiments such as the Pierre Auger Observatory, which covers an area in Argentina roughly the size of Luxembourg (of course with sparse sampling).

Furthermore, while at low energies we can relatively easily identify the particles (they turn put to be mainly atomic nuclei), at the higher energies we have to use very indirect techniques. Combining the results from these many different techniques and experiments is complicated, and this is one major source of disagreement in the field. There is room for disagreement about the fundamental observational facts themselves, even before theoretical interpretation. And once one moves away from the basic observations into interpretation, then further uncertainties are introduced. Relating the observed fluxes of particles to their putative production spectra in sources requires a model for their propagation through the intervening magnetic fields, radiation fields and matter distributions, many of which are poorly known.

Of course some facts are indisputable. That the particles exist is certain. The broad features of the all-particle energy spectrum (where one simply lumps all the particles together and just measures their energy) are by now well established except at the very highest energies. The total power requirements are fairly clear. But the composition at the higher energies is quite uncertain, the origin of some fine structure that has recently been seem in precision measurements at lower energies is unclear, there is a long-standing tension between the source spectra favoured by propagation theory and acceleration theory, we do not have a satisfactory propagation model that fits all the data (especially the anisotropy in arrival directions) etc. As in most areas of science one has a mixture of some well-established facts, but a lot of uncertainty in the detail and especially in the interpretation. The challenge is to synthesise a coherent scheme that is also consistent with the rest of astrophysics and physics. Depending on the relative weight one attaches to different pieces of evidence and theoretical preferences, it is possible to have legitimate peer disagreement on many aspects while still agreeing on certain basics.